# Depth-resolved resonant inelastic x-ray scattering at a superconductor/half-metallic ferromagnet interface through standing-wave excitation

Cheng-Tai Kuo[1,2,*], Shih-Chieh Lin[1,2], Giacomo Ghiringhelli[3], Yingying Peng[3,#], Gabriella Maria De Luca[4], Daniele Di Castro[5], Davide Betto[6], Mathias Gehlmann[1,2], Tom Wijnands[7], Mark Huijben[7], Julia Meyer-Ilse[2], Eric Gullikson[2], Jeffrey B. Kortright[2], Arturas Vailionis[8], Nicolas Gauquelin[7,9], Johan Verbeeck[9], Timm Gerber[1,2,10], Giuseppe Balestrino[5], Nicholas B. Brookes[6], Lucio Braicovich[3], Charles S. Fadley[1,2,†]

[1] *Department of Physics, University of California Davis, Davis, California 95616, USA*
[2] *Materials Sciences Division, Lawrence Berkeley National Laboratory, Berkeley, California 94720, USA*
[3] *CNR-SPIN and Dipartimento di Fisica Politecnico di Milano, Piazza Leonardo da Vinci 32, Milano I-20133, Italy.*
[4] *Dipartimento di Fisica Universita` di Napoli 'Federico II' and CNR-SPIN, Complesso Universitario di Monte Sant'Angelo, via Cinthia, Napoli I-80126, Italy*
[5] *CNR-SPIN and Dipartimento di Ingegneria Civile e Ingegneria Informatica, Università di Roma Tor Vergata, Via del Politecnico 1, I-00133 Roma, Italy*
[6] *European Synchrotron Radiation Facility, 6 rue Jules Horowitz, B.P. 220, Grenoble Cedex F-38043, France*
[7] *Faculty of Science and Technology and MESA+ Institute for Nanotechnology, University of Twente, Enschede 7500 AE, The Netherlands*
[8] *Geballe Laboratory for Advanced Materials, Stanford University, Stanford, California 94305, USA*
[9] *Electron microscopy for materials Science (EMAT), University of Antwerp, Groenenborgerlaan 171, B-2020 Antwerp, Belgium*
[10] *Peter Grünberg Institut PGI-6, Research Center Jülich, 52425 Jülich, Germany*
[#] Present address: Department of Physics and Seitz Materials Research Laboratory, University of Illinois, Urbana, IL 61801, USA

[*] Corresponding author: chengtaikuo@lbl.gov
[†] Corresponding author: fadley@lbl.gov

## ABSTRACT

We demonstrate that combining standing-wave (SW) excitation with resonant inelastic x-ray scattering (RIXS) can lead to depth resolution and interface sensitivity for studying orbital and magnetic excitations in correlated oxide heterostructures. SW-RIXS has been applied to multilayer heterostructures consisting of a superconductor $La_{1.85}Sr_{0.15}CuO_4$ (LSCO) and a half-metallic ferromagnet $La_{0.67}Sr_{0.33}MnO_3$ (LSMO). Easily observable SW effects on the RIXS excitations were found in these LSCO/LSMO multilayers. In addition, we observe different depth distribution of the RIXS excitations. The magnetic excitations are found to arise from the LSCO/LSMO interfaces, and there is also a suggestion that one of the *dd* excitations comes from the interfaces. SW-RIXS measurements of correlated-oxide and other multilayer heterostructures should provide unique layer-resolved insights concerning their orbital and magnetic excitations, as well as a challenge for RIXS theory to specifically deal with interface effects.

**TEXT**

Resonant inelastic x-ray scattering (RIXS) is a photon-in/photon-out synchrotron-based spectroscopy that has been shown to uniquely probe the charge transfer, *dd*, magnetic, phonon and other excitations in correlated oxides and other systems, and has been extensively reviewed elsewhere [1,2]. RIXS is considered to be a probe of bulk properties, at depths of the order of 1000 Å, although in fact, the penetration and escape depths of the resonant x-rays can be significantly reduced for excitations at a strong absorption edge of a majority elemental constituent [3], and thus the actual sensing depth is somehow ill defined and variable from sample to sample. It would thus be desirable to give RIXS more quantitative depth sensitivity, for example to investigate interfaces in oxide heterostructures, which are known to show emergent properties (e.g. 2D electron gases, interface-induced ferromagnetism) not present in the single constituents [4] with these triggering intense interest and many publications on various oxide interfaces [5]. We here demonstrate that, by using standing-wave (SW) excitation from multilayer heterostructures, interface-specific RIXS information can be achieved.

It is well known that a strong Bragg reflection from a multilayer heterostructure or a single crystal creates a SW inside and above the sample, and that it can be used to excite x-ray or photoelectron emission with resulting depth resolution [6,7,8,9,10,11]. Prior reviews of these developments using multilayer reflection from members of our group provide additional background [12,13,14,15], including a detailed discussion of the x-ray optical theoretical modeling program that we will use to interpret our data: Yang X-ray Optics (YXRO) [3]. The relevant Bragg equation is $n\lambda_x = 2d_{ML}\sin\theta_{inc}$, where *n* is the order of the reflection, $\lambda_x$ the x-ray wavelength, $d_{ML}$ the bilayer repeat spacing in the multilayer and $\theta_{inc}$ the incidence angle relative to the multilayer. It is simple to show that, for first-order Bragg reflection, the period of the SW electric-field intensity $|E^2| \equiv \lambda_{SW} = d_{ML}$, where $\lambda_{SW}$ is the wavelength of the SW vertical to the layers and the interfaces between them. The SW can be swept through the sample in two principal ways: scanning the incidence angle $\theta_{inc}$ over the Bragg reflection through a rocking curve (the method used here), and scanning the photon energy, i.e., the photon wavelength $\lambda_x$ through the Bragg reflection. When spanning the whole Bragg peak, both methods shift the SW spatially by one half of its period in a direction perpendicular to the interfaces in the multilayer. The standard formula for the SW intensity at a given depth $z$ below the surface is:

$$I(\theta_{inc}) \propto 1 + R(\theta_{inc}) + 2\sqrt{R(\theta_{inc})}\, f \cos[\varphi(\theta_{inc}) - 2\pi(z/\lambda_{SW})] \quad , \quad (1)$$

where $R(\theta_{inc})$ is the reflectivity, $f$ the fraction of atoms in coherent positions for Bragg reflection, $\varphi(\theta_{inc})$ the phase difference between incident and scattered waves, and $z/\lambda_{SW}$ the vertical position of a given layer or interface of interest, as normalized to the SW period. The third term here represents the SW modulation. Although the basic physics of the SW formation is contained in Equation (1), the YXRO program actually calculates the SW in a more accurate way, including x-ray attenuation and multiple scattering or dynamical diffraction effects [3].

In this work, we show that SW excitation in RIXS can be used to provide enhanced depth and interface sensitivity to the technique. We have chosen to probe the interface between the superconducting cuprate $La_{1.85}Sr_{0.15}CuO_4$ (LSCO) and the half-metallic ferromagnetic manganite $La_{0.67}Sr_{0.33}MnO_3$ (LSMO) in an assessment of SW-RIXS capabilities. In this cuprate/manganite heterostructure, De Luca *et al.* found a strong charge transfer from Mn to Cu ions using electron energy loss spectroscopy and x-ray circular dichroism [16]. The interfacial $CuO_2$ planes of the cuprate develop weak ferromagnetism associated with the charge transfer from the $MnO_2$ planes of the manganite, and the Dzyaloshinskii-Moriya interaction propagates the magnetization from the interfacial $CuO_2$ planes into the superconductor, leading to a depression of its superconducting critical temperature. Information on the length scale of this charge transfer at the LSCO/LSMO interface and its relationship to the *dd* and magnetic excitations could provide a more complete understanding of this interface coupling, with LSCO/LSMO thus providing an ideal system for testing the depth resolution of SW-RIXS.

Multilayers of $(LSCO_n/LSMO_m)_p$ ($n$= 2 unit cell (uc), $m$= 7 uc, and $p$ = 20 repeats) were grown by pulsed laser deposition, either on SrO-terminated and on $TiO_2$-terminated $SrTiO_3$ (STO) substrates, *in situ* controlled by reflection high-energy electron diffraction. The details of the growth of the LSCO/LSMO heterostructures can be found elsewhere [16,17]. The individual layers are thus nominally LSCO = 26.4 Å and LSMO = 27.0 Å, based on bulk properties, yielding an estimated $d_{ML}$ of 53.4 Å. More precise measurements of these dimensions using scanning transmission electron microscopy, together with high-angle annular dark field imaging (STEM-HAADF) and electron energy loss spectroscopy (EELS) were performed on a Titan 80-300 microscope equipped with an aberration corrector for the probe forming lens used at 300 kV acceleration voltage with a 20 mrad convergence angle and a collection angle of 40-95 mrad for HAADF imaging. EELS was used to determine the chemistry at each LSMO(top)/LSCO(bottom) and LSCO(top)/LSMO(bottom) interface which always consist of the sequence -$La_{0.9}Sr_{0.1}O$-$La_{0.9}Sr_{0.1}O$-$CuO_2$-$La_{0.66-x}Sr_{0.33+x}O$-$MnO_2$-$La_{0.66}Sr_{0.33}O$- and -$La_{0.66}Sr_{0.33}O$-$MnO_2$-$La_{0.9-x}Sr_{0.1+x}O$-$CuO_2$-$La_{0.9}Sr_{0.1}O$-$La_{0.9}Sr_{0.1}O$- respectively (0<x<0.15). La/Sr ratios are subject to a 5% error inherent to the measurement method. One aspect of this data is shown in Figures 1(d) and 1(e), in which the $TiO_2$-termination is shown to be less regular as a multilayer. Therefore, we present in the main text the results on the SrO-terminated multilayer, which has superior structural regularity, and discuss in detail the TiO$_2$-terminated case in our Supplemental Material [18], because the SW effects on RIXS were a more complex to analyze due to irregularities in its bilayer spacings as seen in STEM images. The SW-RIXS measurements on both samples were performed at ID32 of ESRF using the high-resolution ERIXS spectrometer [19]. The total instrumental energy resolution was set at 70 meV, determined as the FWHM of the non-resonant diffuse scattering from silver paint adjacent to the sample. The multilayer samples were cooled down to ~20K by liquid He, and thus below the superconducting T$_c$ of bulk LSCO (~30-40K), and the ferromagnetic T$_c$ of bulk LSMO (~270-298 K). The RIXS data were collected near the Cu L$_3$ edge.

Given the multilayered structure of the sample, as shown in Figure 1(a), we can choose the incidence angle $\theta_{inc}$ to match the Bragg conditions near the Cu L$_3$ edge ($hv$ = 931.2 eV) for the sample period $d_{ML} \approx$ 53.4 Å. From the measured imaginary part of the index of refraction for the multilayer (see Supplemental Material [18]), we estimate the effective exponential decay length of the x-ray intensity, including incidence and exit, to be about $\Lambda_{x,eff} \approx$ 54 Å, which is drastically lower than the ~1000 Å that are often assumed in the literature, due to the strong absorption resonance. Coincidentally, the decay length approximately matches the bilayer period, which means that the RIXS signal is almost completely attenuated at the bottom of the multilayer at a depth of 20 periods or ~1070 Å. Indeed, 95% of the RIXS signal arise from a depth of $3\Lambda_{x,eff} \approx$ 162 Å or about the three topmost bilayers. As noted above, for first-order Bragg reflection, the SW period $\lambda_{SW} = d_{ML}$, and by scanning $\theta_{inc}$ in the vicinity of the nominal Bragg position, the maxima of the SW moves by $\lambda_{SW}/2 \approx$ 27 Å across the interface. Other details concerning the characterization of the sample grown on SrO-terminated STO, as well as the second one grown on TiO$_2$-terminated STO, are presented in Supplemental Material [18].

The intensities of the individual RIXS excitations as a function of incidence angle, which we call rocking curves (RCs), are thus modulated by the moving SW field, schematically shown in Figures 1(a) and 1(b). Figure 1(b) shows the scattering geometry in real space and momentum space. The incident beam hits the sample at an angle $\theta_{inc} \approx 7°$ from the surface and is reflected by the multilayer with a Bragg vector $\mathbf{q}_{SW}$ normal to the surface; the RIXS signal is collected in backscattering at $\theta_{scatt} \approx 30°$, resulting in a RIXS scattering vector $\mathbf{q}_{RIXS}$ mostly parallel to the surface. Throughout the RC, which means with increasing $\theta_{inc}$ and $q_{SW}$, the standing wave develops initially in the low absorption LSMO layer and shifts by $d_{ML}/2$ into the LSCO layer as the multilayer Bragg peak is crossed. Figure 1(c) shows a representative Cu $L_3$ edge RIXS spectrum from the SrO-terminated LSCO/LSMO multilayer, and it is clear that quasi-elastic, magnetic, and *dd* excitations are observed. The RIXS spectrum in the range of 0 to 500 meV consists of the elastic peak, phonon excitations, and magnetic (mainly single magnon and bimagnon) excitations [20,21,22,23,24,25]. The bimagnon signal in RIXS results from the sudden change of the superexchange magnetic interaction in the intermediate state [26,27]. The spectral range from 1 to 2.5 eV is dominated by *dd* excitations [2,21,22,23,28], which are partly resolved into a doublet and a low-energy shoulder whose assignment was already discussed in Ref. 29. We here focus on the RCs of the *dd*, magnetic, and quasi-elastic (elastic + phonon) RIXS excitations from the SrO-terminated multilayer, although noting that, at our resolutions, cleanly separating them all by peak fitting must be done carefully to avoid artifacts.

First, we discuss the RCs of the *dd* excitations. The *dd* excitations can be ascribed to the transfer of the 3d hole from the $d_{x2-y2}$ orbital to the $d_{z2}$, $d_{xy}$, and $d_{yz}/d_{xz}$ orbitals [28]. In Fig. 2(a) the *dd* excitations are deconvoluted by peak fitting into the $d_{z2}$, $d_{xy}$, and $d_{yz}/d_{xz}$ components. In order to observe the SW movement across the interfaces, the RIXS *dd* excitation spectra were collected while varying the incidence angle between 6.5° and 10°, thus yielding three RCs shown in Figure 2(c). All the experimental and theoretical RCs are normalized to a maximum of unity and are offset vertically for readability. The fractional modulation of each RC can thus be read directly from the ordinate scale. The intensity of all *dd* excitations is modulated by 15-20%, meaning that the SW has a clear influence on the RIXS process: this is the experimental demonstration that SW-RIXS is feasible. We note also that these three RCs show a very similar shape, with intensity minima at ~8.2°, thus indicating a very similar depth distribution. This is not surprising, as the cross section of the dd excitations is not expected to depend on the details of the local coordination of the $Cu^{2+}$ ions; at most their energy might change from the surface to the bulk layers, but in this experiment we did not attempt to detect those energy shifts, as these are expected to be small. One can argue that normalization to the "flat" wings of an RC for which reflectivity and SW modulation are minimal is a better choice, but it can be more difficult to do if instrumental effects such as beam movement along the sample or slight changes in self-absorption or excitation cross section during a scan lead to a complex, sloping background. Our normalization choice should not affect any of our conclusions, however. We illustrate this in Figures 3(d)-(e), where we show the measured and calculated reflectivity, and its second derivative. It seems clear that no significant SW effects exist at the edges of the 7.0-9.5° angle of our experimental RCs (Figs. 2(c),(d)) and calculated RCs (Fig. 3(c)).

As the RC intensity modulation is significant, we now try to relate these RCs to an approximate depth distribution of the loss processes involved, by simulating the RIXS process using the previously mentioned YXRO program [3]. Two key inputs to this program are the resonant index of refraction and the detailed structure (e.g. thickness of individual layers) of the sample. The resonant index of refraction has been derived by measuring the multilayer x-ray absorption curve and using Kramers-Kronig analysis (see Supplemental Material [18]). Note that all of the resonant Cu atoms are assumed to be uniform in the calculations. It is possible that the Cu atoms near the top and bottom interfaces have different environments (e.g. position distortions, charge transfer, etc.). This could lead to the difference in the x-ray absorption and a slight change in the simulated SW electric-field distribution, but it will not change the conclusions of this work. Future SW works on deriving more interface-like x-ray absorption should help improve our understanding of the x-ray optical effects in RIXS.

The thicknesses of the individual LSCO and LSMO layers are determined from high-resolution STEM-HAADF images (see Supplemental Material [18]), and used as inputs for YXRO. In Figure 3, we show various results from these simulations. Figure 3(a) shows the calculated SW electric-field strength $|E(z,\theta_{inc})|^2$ as a function of depth and $\theta_{inc}$, including a layer of $CO_x$-containing surface contaminants. This plot illustrates the scan of the SW vertically in the sample, and makes it clear that the SW has the principal effect of enhancing the RIXS signal from the first LSCO-top/LSMO-bottom interface over the lower lying LSCO layers and interfaces. Figure 3(b) shows the model structure on which simulations have been carried, focusing in particular on the first interface. The simulated RCs arising from these different regions are shown in Figure 3(c). It is clear that the calculated RCs for the different depth ($\Delta z$) are markedly different. For example, the RC from the LSMO-top/LSCO-bottom interface ($\Delta z$ = 0 Å) has a minimum at ~ 7.9°, while that from the LSCO-top/LSMO-bottom interface ($\Delta z$ = 22.5 Å) has a minimum at 8.5° to 9.0°. We determine depth distribution of each RIXS by comparing its experimental RC to a weighted sum of these depth-resolved RCs (depth step of 2.5 Å) until the best fit to the experimental results is found. This has been done both using least-squares fittings and visual inspection of the calculated RCs to the experimental data. Comparing these simulations to the dd experimental data in Figure 2(c), we find that the experimental RCs match the average of the RCs from the whole LSCO layer in Figure 3(c), as shown by the solid curves. We can thus conclude that all three *dd* excitations show very similar behavior, with profiles suggesting that this part of the RIXS spectrum is quite independent from the position inside the LSCO layer, as indicated by the inset in Figure 2(c).

We now consider the RCs of the quasi-elastic and magnetic excitations, as shown in Figure 2(b). The excitations in this range are more complex to analyze since the magnetic excitations lie very close to the phonon peaks and the elastic zero-loss line, and are also relatively weak. The quasi-elastic peak includes the elastic zero-loss line and phonon excitations [19,20,21,23]. To avoid spurious intensity variations in the fittings, we thus report in Figure 2(d) RCs as more statistically accurate sums over the peak fitting groups in Figure 2(b), that is, over elastic + phonon + biphonon and over magnon + bimagnon. For the quasi-elastic RC that shows minima at ~8.4°, the depth distribution agrees with the calculated curves which have their origin over most of the LSCO layer (20 Å), excluding the top interface region (see the bottom inset in Figure 2(d)). The RCs of the summed magnetic excitations show a similar behavior as the *dd* excitations, but with smaller intensity modulation (~8%) and minimum at 8.2°. Again, we compare the experimental RC of magnetic excitations to a weighted sum of the simulations in Figure 3(c) to determine its depth distribution, and this yields the conclusion of a depth distribution peaked at the LSMO-top/LSCO-bottom and LSCO-bottom/LSMO-top interface. We have carried out various simulations by summing over the depth-resolved RC curves to compare with the experimental data, which includes summing over the whole LSCO layer, summing from the bottom LSCO interface, summing from the top LSCO interface, and summing over from both top and bottom LSCO/LSMO interfaces. The experimental RC of magnetic excitations agrees best with the sum of the calculated top-8-Å and bottom-8-Å curves: we interpret this as an enhancement of the magnetic signal at the interfaces as sketched in the top inset in Figure 2(d).

To have more quantitative depth profiles of various excitations, these experimental RCs are fit by Equation (2)

$$I_{RC,j}^{Expt}(\theta_{inc,k}) = \sum_{z_i} W_{ji}(z_i,\theta_{inc,k}) I_{RC,j}^{Calc}(z_i,\theta_{inc,k}) \exp(-z_i/\Lambda_{x,eff}), \qquad (2)$$

where $I_{RC,j}^{Expt}(\theta_{inc,k})$ is an experimental RC at incidence angle (with j = magnetic or quasi-elastic, for example), $I_{RC,j}^{Calc}(z_i,\theta_{inc,k})\exp(-z_i/\Lambda_{x,eff})$ is one of the calculated RCs in Fig. 3(c) below, and $W_{ji}(z_i,\theta_{inc,k})$ is a weighting coefficient in a fitting procedure that we have derived using the quasi-Newtonian Broyden, Fletcher, Goldfarb, and Shanno (BFGS) method [30]. Finally plotting the summed amplitudes of these weighting factors at a given $z_i$ interval of 2.5 Å as $W_{ji}^{Sum}(z_i) = \sum_{\theta_{inc,k}} W_{ij}(z_i,\theta_{inc,k})$ then yields a quantitative estimate of the depth distributions. For example, the magnetic excitations in Fig. 2(e) are found to occupy about 3 Å near the LSMO-top/LSCO-bottom interface, with a weaker contribution also from the next LSCO-top/LSMO-bottom interface. The quasi-elastic excitation in Fig. 2(f) is found to show contributions from the full LSCO layer, although weighted away from the top interface. The results agree with the more qualitative fitting described above.

This result that shows the depth distribution of magnetic excitations mainly originate from 3-8 Å interfacial regions in LSCO is far from trivial. We note that the magnon excitations seen in RIXS correspond to damped spin-waves from the 2D antiferromagnetic lattice in the $CuO_2$ planes. Upon hole doping the magnon energy is unchanged but the damping grows and the bimagnon contributions is progressively washed out [31], therefore a stronger overall magnetic

RIXS intensity at the interfaces might be explained by the charge transfer from LSMO to LSCO that locally reduces the hole doping of the cuprate. This result complements what had been found by studying the x-ray absorption spectra and the magnetic circular dichroism of these LSCO/LSMO interfaces, that a weak ferromagnetic order is induced in the cuprate by the manganite: we conclude thus that the latter is not reducing but, on the contrary, enhances the antiferromagnetic short range correlation of the cuprate.

As a final aspect of the experimental data, in Supplemental Material [18] we discuss analogous SW-RIXS results for the structurally less well-defined multilayer grown on $TiO_2$-terminated STO. These include complementary SW photoemission (SW-XPS) measurements at exactly the same photon energy. Although the stacking sequence of bilayers is not regular in this sample, and this strongly influences the SW form, the SW-RIXS results are in qualitative agreement with those for SrO-terminated growth, but also suggest that the $d_{z2}$ $dd$ excitation is slightly enhanced at the LSCO-top/LSMO-bottom interface, possibly signaling a local modification of the crystal field, ie of the $Cu^{2+}$ ion coordination. The SW-XPS RCs for Cu 3p and Mn 3p RCs (Figure S9) are found to be well predicted by YXRO calculations for the best-fit geometry. Thus, these additional SW-RIXS and SW-XPS results for a less ideal sample configuration further confirm our analysis of the SW-RIXS data for SrO-terminated growth, and might have some more indication within the $dd$ excitations, but needs a better sample in the future to confirm this.

In conclusion, we have demonstrated that soft x-ray RIXS is sensitive to standing waves. For the LSCO/LSMO multilayer heterostructures, thanks to advanced x-ray optical theoretical simulations, we could interpret qualitatively the experimental results in terms of relative enhancement of some of the excitations at the interfaces and with respect to the bulk regions of LSCO. In particular we found that for the sample grown on an SrO-terminated STO substrate the magnetic excitations have their origin from both the top and bottom LSCO/LSMO interfaces. Future studies with superlattices of more ideal geometry should permit more quantitatively determining RIXS depth distributions, including differences in the *dd* excitations. Applying SW-RIXS to quasi-2D quantum materials (e.g. topological insulators and transition-metal dichalcogenides) is also promising, with the SW in these systems resulting from Bragg reflection from different crystal planes, and RIXS thus in principle being given atomic-layer sensitivity. Although there are at present no theoretical simulations of RIXS that take account of the depth of excitations, we suggest that future measurements of this type on more regular sample configurations will stimulate them, and that SW-RIXS will open up a new spatial dimension to this already powerful technique.

**ACKNOWLEDGMENTS**

This work was supported by the US Department of Energy under Contract No. DE-AC02-05CH11231 (Advanced Light Source), and by DOE Contract No. DE-SC0014697 through the University of California Davis (salary and travel expenses for C.-T.K., and S.C.L. and C.S.F.). C.S.F. has also been supported for salary by the Director, Office of Science, Office of Basic Energy Sciences (BSE), Materials Sciences and Engineering (MSE) Division, of the U.S. Department of Energy under Contract No. DE-AC02-05CH11231, through the Laboratory Directed Research and Development Program of Lawrence Berkeley National Laboratory, through a DOE BES MSE grant at the University of California Davis from the X-Ray Scattering Program under Contract DE-SC0014697, through the APTCOM Project, "Laboratoire d'Excellence Physics Atom Light Matter" (LabEx PALM) overseen by the French National Research Agency (ANR) as part of the "Investissements d'Avenir" program, and from the Jűlich Research Center, Peter Grűnberg Institute, PGI-6. J.V. and N.G. acknowledge funding through the GOA project "Solarpaint" of the University of Antwerp. The microscope used in this work was partly funded by the Hercules Fund from the Flemish Government. The RIXS experiment was made the beam line ID32 of the ESRF using the ERIXS spectrometer. GG and YYP were supported by ERC-P-ReXS project (2016-0790) of the Fondazione CARIPLO and Regione Lombardia, in Italy.

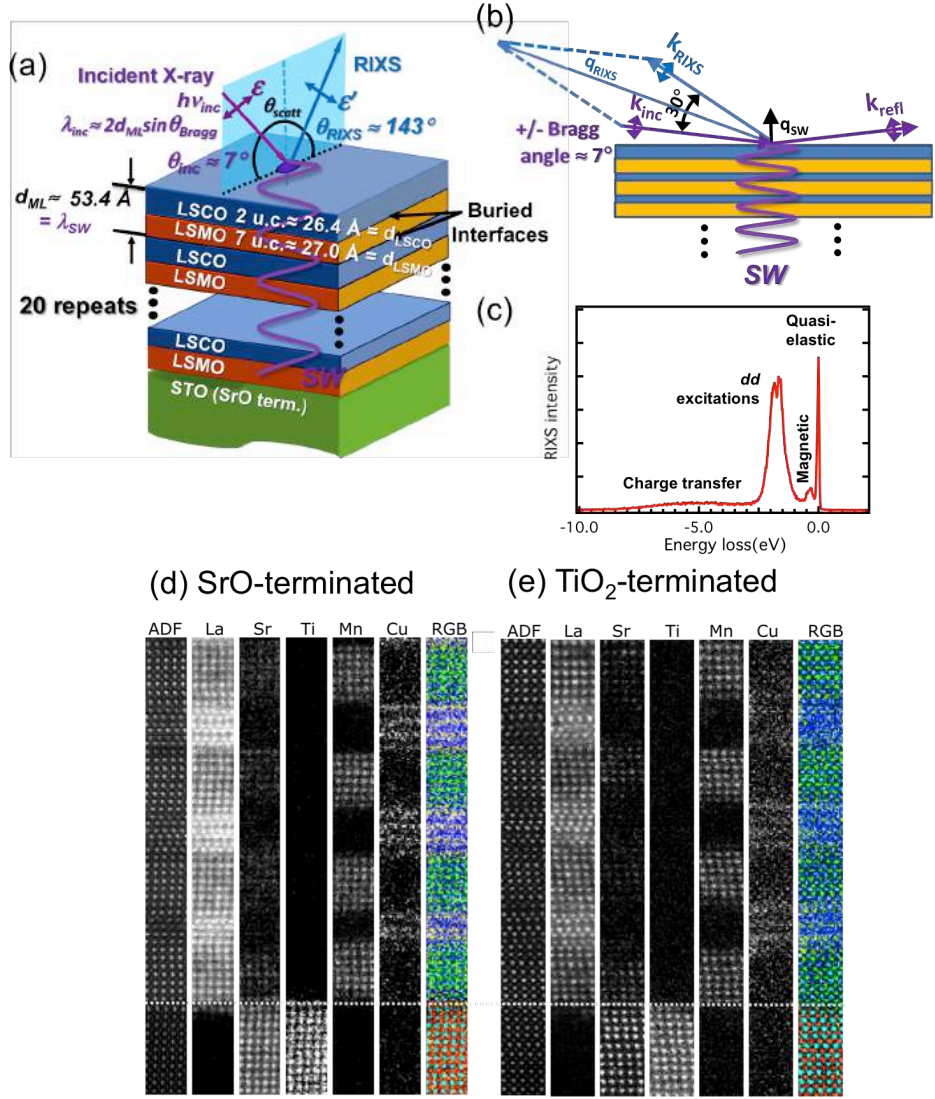

**FIG. 1.** Schematic illustrations of the standing-wave (SW) excited resonant inelastic x-ray scattering (RIXS) measurement. (a) Diagram of the multilayer sample with bilayer period $d_{ML}$, including the geometry of the exciting x-ray beam, the scattered photons and the standing wave indicated. The multilayer samples consist of 20 bilayers of 2 unit cells of $La_{1.85}Sr_{0.15}CuO_4$ (LSCO) and 7 unit cells of $La_{0.67}Sr_{0.33}MnO_3$ (LSMO), grown epitaxially on SrO-terminated STO substrate. The dimensions shown are nominal, based on bulk lattice parameters. (b) The SW-RIXS experimental geometry in real space and momentum space. (c) A typical RIXS spectrum, from the SrO-terminated growth, that exhibits quasi-elastic, magnetic, and *dd* excitations. (d) and (e) The STEM-HAADF and EELS results near the initial growth on STO for both the SrO-terminated growth and the less regular TiO$_2$-terminated. In the RBG images Ti is orange, Mn is green, Cu is blue, Sr is turquoise, La is green.

SW-RIXS--Growth on SrO-terminated STO:

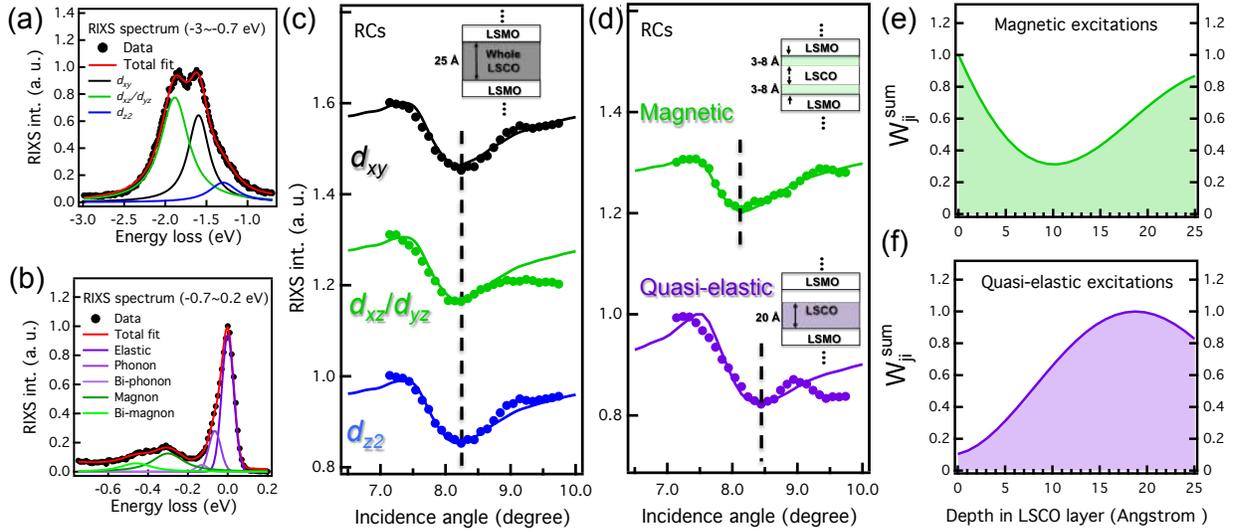

**FIG. 2.** SW-RIXS of various excitations. RIXS spectra of (a) *dd* excitations and (b) quasi-elastic and magnetic excitations. The *dd* excitations have three components: $d_{xy}$, $d_{xz}/d_{yz}$, and $d_{z2}$. The quasi-elastic intensity includes three components, the zero-loss or elastic line and phonon excitations, with these three components being summed to give the rocking curve (RC). The magnetic spectra are fit with two components (magnon and bi-magnon), whose intensities are summed to yield the magnetic RC. (c) The experimental *dd*-excitations RCs (data points) together with YXRO calculations (lines). (d) The experimental RCs for the magnetic and quasi-elastic excitations (data points) together with YXRO calculations (lines). (e) The summed weighting factors from Equation (2) for the magnetic excitations. (f) The summed weighting factors from Equation (2) for the quasi-elastic excitations.

SW-RIXS—Simulations for growth on SrO-terminated STO:

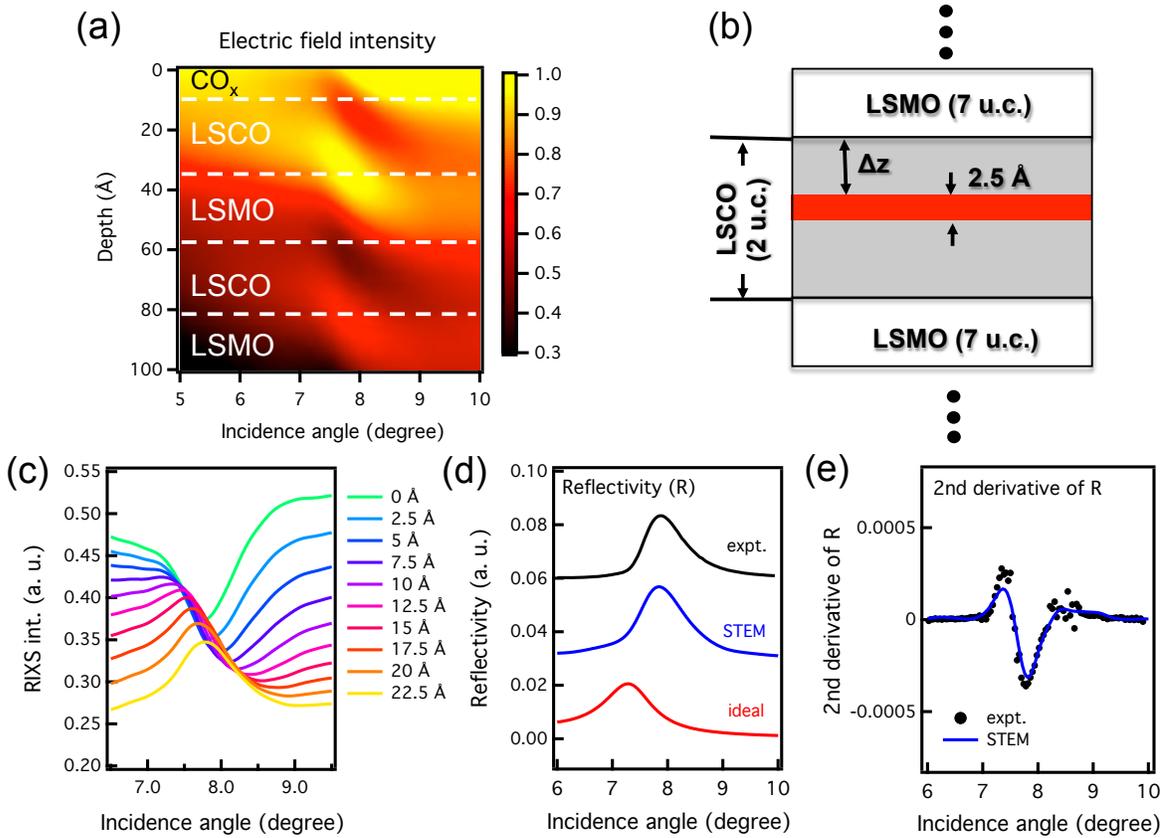

**FIG. 3.** (a)-(c) Results of x-ray optical simulations of the SW effects and the depth-resolved RCs, for growth on SrO-terminated STO. (a) The depth and incidence-angle dependence of the SW electric-field intensity $|E(z,\theta_{inc})|^2$. Note in particular the movement of the SW through the top two LSCO-top/LSMO-bottom interfaces. (b) The model sample profile of LSCO that is used to simulate the RCs resulting from from 0 to 22.5 Å, with "delta-layer" thickness of 2.5 Å. (c) The calculated RCs for various Δz values. (d) Experimental (black curve) and calculated (blue and red curves) soft x-ray reflectivity. (e) Experimental (black dots) and calculated 2$^{nd}$ derivative of the reflectivity data. The blue curves in (d)(e) are for the STEM-determined sample configuration, allowing fully for the non-uniformity of some bilayer thicknesses and the red curve in (d) is for the ideal sample configuration. (see Supplemental Material [18]).

# Supplemental Material

# Depth-resolved resonant inelastic x-ray scattering at a superconductor/half-metallic ferromagnet interface through standing-wave excitation

Cheng-Tai Kuo[1,2,*], Shih-Chieh Lin[1,2], Giacomo Ghiringhelli[3], Yingying Peng[3,#], Gabriella Maria De Luca[4], Daniele Di Castro[5], Davide Betto[6], Mathias Gehlmann[1,2], Tom Wijnands[7], Mark Huijben[7], Julia Meyer-Ilse[2], Eric Gullikson[2], Jeffrey B. Kortright[2], Arturas Vailionis[8], Nicolas Gauquelin[7,9], Johan Verbeeck[9], Timm Gerber[1,2,10], Giuseppe Balestrino[5], Nicholas B. Brookes[6], Lucio Braicovich[3], Charles S. Fadley[1,2,†]

**Affiliations**

[1] *Department of Physics, University of California Davis, Davis, California 95616, USA*
[2] *Materials Sciences Division, Lawrence Berkeley National Laboratory, Berkeley, California 94720, USA*
[3] *CNR-SPIN and Dipartimento di Fisica Politecnico di Milano, Piazza Leonardo da Vinci 32, Milano I-20133, Italy.*
[4] *"Dipartimento di Fisica "E. Pancini" Universita` di Napoli 'Federico II' and CNR-SPIN, Complesso Universitario di Monte Sant'Angelo, Napoli I-80126, Italy*
[5] *CNR-SPIN and Dipartimento di Ingegneria Civile e Ingegneria Informatica, Università di Roma Tor Vergata, Via del Politecnico 1, I-00133 Roma, Italy*
[6] *European Synchrotron Radiation Facility, 6 rue Jules Horowitz, B.P. 220, Grenoble Cedex F-38043, France*
[7] *Faculty of Science and Technology and MESA+ Institute for Nanotechnology, University of Twente, Enschede 7500 AE, The Netherlands*
[8] *Geballe Laboratory for Advanced Materials, Stanford University, Stanford, California 94305, USA*
[9] *Electron microscopy for materials Science (EMAT), University of Antwerp, Groenenborgerlaan 171, B-2020 Antwerp, Belgium*
[10] *Peter Grünberg Institut PGI-6, Research Center Jülich, 52425 Jülich, Germany*
[#] *Present address: Department of Physics and Seitz Materials Research Laboratory, University of Illinois, Urbana, IL 61801, USA*

[*] Corresponding author: chengtaikuo@lbl.gov
[†] Corresponding author: fadley@lbl.gov

## ➢ Introduction:

In our Supplemental Material, we present detailed x-ray optical and structural characterization of the two samples studied: $La_{1.85}Sr_{0.15}CuO_4/La_{0.67}Sr_{0.33}MnO_3$ (LSCO/LSMO) multilayers grown on SrO-terminated $SrTiO_3$ (STO)--as discussed in detail in the main text, and $TiO_2$-terminated STO--a less regular multilayer that nonetheless exhibits standing-wave excited inelastic resonant x-ray scattering (SW-RIXS) results that qualitatively support the conclusions in the main text, but at the same time illustrates SW fine structure that can complicate the interpretation of the results.

## ➤ Resonant index of refraction of LSCO:

The resonant index of refraction is one of the key inputs for x-ray optical simulations in order to generate the suitable reflectivity and RIXS rocking curves (RCs). Figure S1(a) shows the Cu $L_3$ edge x-ray absorption (XAS) spectrum measured from our LSCO/LSMO multilayers. The resonant index of refraction was derived using Kramers-Kronig analysis from the experimental XAS data [S1], and is shown in Figure S1(b), including the real part (delta) and imaginary part (beta) of LSCO around the $L_3$ resonant edge. For our experiments, the photon energy was tuned to 931.2 eV, thus both maximizing the reflectivity and SW strength as well as yielding a strong RIXS signal.

X-ray optical properties--Growth on SrO-terminated STO:

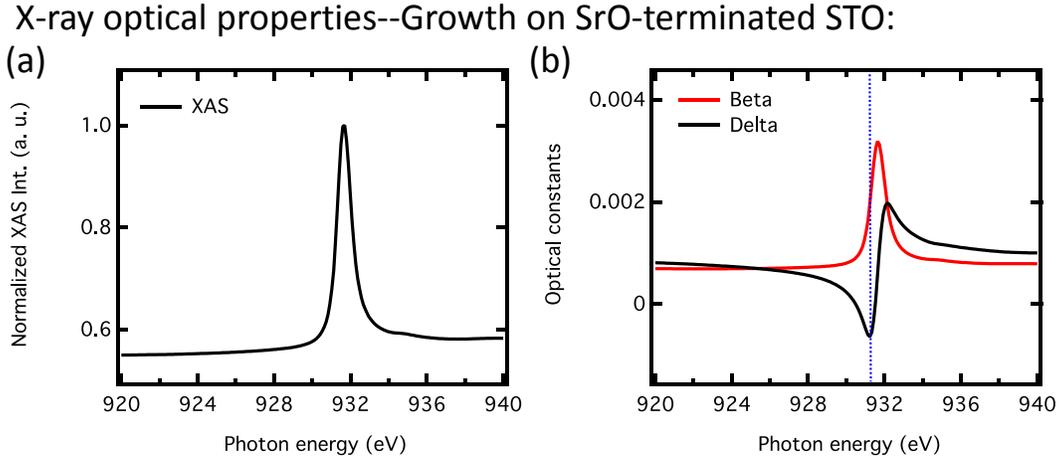

**FIG. S1.** (a) X-ray absorption (XAS) spectrum of the LSCO/LSMO multilayer grown on SrO-terminated STO and (b) its corresponding resonant index of refraction (delta and beta) derived from Kramers-Kronig transformation. The vertical line indicates the energy of our experiments: 931.2 eV.

## ➤ The attenuation length in RIXS:

In the limit appropriate to our sample of weak reflectivity ($R(\theta) < 0.03$), the attenuation length of x-ray emission ($\Lambda_x$) can be calculated from the imaginary part (beta) of the index of refraction for the multilayer through $\Lambda_x = \lambda_x/(4\pi\beta)$. The decay length of the x-ray intensity, including incidence and exit direction of x-ray, is defined as the effective attenuation length ($\Lambda_{x,eff}$) and it can be calculated from equation (S1). We here neglect the attenuation due to the LSMO due to its non-resonant and much greater $\Lambda_x$ of ~1900 Å.

$$\frac{1}{\Lambda_{x,eff}} = \frac{1}{\Lambda_x \sin\theta_{inc}} + \frac{1}{\Lambda_x \sin\theta_{exit}} , \quad (S1)$$

where $\theta_{inc}$ is the incidence angle of the incoming x-ray and $\theta_{exit}$ is the exit angle of the outgoing x-ray. For our case, and allowing for resonant excitation at hv = 931.2 eV, $\Lambda_x$ = 540 Å, $\theta_{inc}\approx7°$, $\theta_{ext}\approx37°$, we finally arrive at $\Lambda_{x,eff} \approx 54$ Å, as discussed in the main text. Thus, the depth from which 95% of intensity will arise will be about 3 $\Lambda_{x,eff} \approx 162$ Å or about 3 bilayers.

## ➢ LSCO/LSMO multilayer grown on SrO-terminated STO--Structural information:
### • STEM, x-ray diffraction and hard x-ray reflectivity

In addition to the resonant index of refraction, the sample geometry is the crucial input for x-ray optical simulations of our experimental results using the YXRO program [S2]. Figure S2(a) shows a scanning transmission electron microscopy with high-angle annular dark field (STEM-HAADF) image with is the result of the alignment of 20 consecutive fast-acquired STEM images [S3] for the LSCO/LSMO multilayer grown on a SrO-terrminated STO. Atomically resolved STEM electron energy loss spectra (EELS-not shown here) was used to determine the exact chemistry at the interfaces, each LSMO(top)/LSCO(bottom) and LSCO(top)/LSMO(bottom) interfaces which always consist of the sequence -$La_{0.9}Sr_{0.1}$O-$La_{0.9}Sr_{0.1}$O-$CuO_2$-$La_{0.66-x}Sr_{0.33+x}$O-$MnO_2$-$La_{0.66}Sr_{0.33}$O- and -$La_{0.66}Sr_{0.33}$O-$MnO_2$-$La_{0.9-x}Sr_{0.1+x}$O-$CuO_2$-$La_{0.9}Sr_{0.1}$O-$La_{0.9}Sr_{0.1}$O- respectively (0<x<0.15). La/Sr ratio are subject to a 5% error inherent to the measurement method. The bilayer stacking and thicknesses are quite regular for this sample. The laterally-averaged thicknesses of the individual LSCO and LSMO layers were determined from this STEM-HAADF image (see the red lines between the layers) by deriving crossover points in the images. The resulting layer-by-layer STEM-based structure was input in the YXRO simulations to generate both the hard and soft x-ray reflectivity data shown in Figs. S2(c) and 4(a), as well as the SW-RIXS RCs in Figs. 2 and 3 of the main text. Fig. S2(b) also shows standard Cu Kα x-ray diffraction (XRD) results for angles around the STO (002) reflection for the LSCO/LSMO multilayer, including $N^{th}$-order satellites SL-N associated with the multilayer Bragg reflections. Dynamical XRD calculations including the detailed atomic structure are also shown, which could not be calculated for the layer-by-layer STEM geometry, but were optimized by trial and error to agree optimally with the data for 20 layers divided as: 11 bilayers of 1.6 u.c. of $d_{LSCO}$ = 12.83 Å and 6.8 u.c. of $d_{LSMO}$ = 3.94 Å, which sum to $d_{ML}$ = 47.29 Å, 6 bilayers of 2.0 u.c. of $d_{LSCO}$ = 13.24 Å and 6.5 u.c. of $d_{LSMO}$ = 4.02 Å, which sum to $d_{ML}$ = 52.62 Å, and 3 bilayers of 2.0 u.c. of $d_{LSCO}$ = 13.36 Å and 6.8 u.c. of $d_{LSMO}$ = 4.04 Å, which sum to $d_{ML}$ = 54.19 Å. The splittings of the SL-2 and SL-3 peaks are clearly predicted correctly, and the overall reflectivity curves in general agree well with one another. The red curve in Fig. S2(b) is an XRD calculation for an ideal structure with no variation in layer thicknesses over the multilayer and the structural parameters of it are 20 repetitions and the value of $d_{ML}$ equals the sum of ideal thickness of 2 u.c. LSCO with $d_{LSCO}$ = 26.4 Å, and resulting 7 u.c. LSMO with $d_{LSMO}$ = 27.0 Å, which sum to $d_{ML}$ = 53.4 Å. This demonstrates the high sensitivity of XRD to deviations from the ideal geometry.

The XRD data and its simulations however deals with a broader area of the sample, and thus might not be the best look at the sample structure for the smaller spot that we study in our SW measurements, and with a more precise layer-by-layer determination of $d_{LSCO,j}$, $d_{LSMO,j}$, and $d_{ML,j}$ for each bilayer from the STEM-HAADF image. Our hard- and soft- x-ray calculations, as simulated with the YXRO program below should be much closer to the physics of the SW-RIXS and SW-photoemission (or SW-XPS) data. The XRD-determined averaged thickness of the period $d_{ML}$ is ~51 Å using the angular separations of $N^{th}$-order satellites, with the relevant individual layer thicknesses showing a total value consistent with the averaged over layers $j$ STEM results, which yield $d^{av}_{LSCO}$ = 23.9±2.7 Å, $d^{av}_{LSMO}$ = 27.4±2.5 Å and, in sum, $d^{av}_{ML}$ = 51.3±3.7Å, with error added in quadrature. These XRD and average STEM nos. thus agree well with one another, and are within 4% of the ideal $d_{ML}$ in Figure 1(a) for bulk materials.

Fig. S2(c) now shows the near-total reflection hard x-ray reflectivity (XRR) data near the $1^{st}$ order Bragg reflection of the multilayer. Also shown are simulated reflectivities from the YXRO program, using the detailed layer-by-layer STEM-determined $d_{LSCO,j}$, $d_{LSMO,j}$, and resulting $d_{ML,j}$ values and the ideal structure. There is good agreement between experiment and theory for the STEM structure, with various peak coincidences indicated by dashed lines, and we conclude that the actual sample structure deviates from the ideal sample structure, but not enough to prevent a semi-quantitative analysis of the RIXS RCs for this sample. The red curve for the ideal geometry is again much different from the other two curves, stressing the need for such measurements for any SW study of a multilayer. These STEM results provide the sample geometries for the following soft x-ray reflectivity, as well as RIXS RCs simulations using the YXRO program in the main text.

STEM and x-ray reflectivity--Growth on SrO-terminated STO:

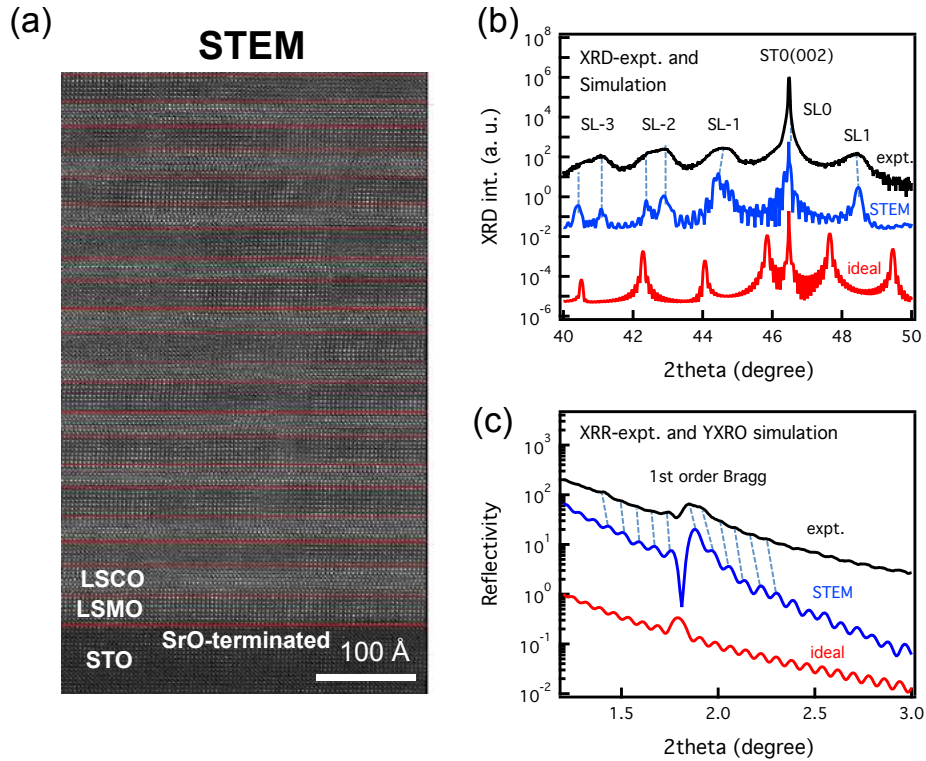

FIG. S2. Structure information for the LSCO/LSMO multilayer grown on the SrO-terminated $SrTiO_3$ substrate. (a) Scanning transmission electron microscopy (STEM) image, with red lines indicating lateral averages to determine the layer thickness variations through the sample. The experimental (b) (002) x-ray diffraction (XRD) patterns compared with dynamical XRD simulations for the STEM structure and the ideal structure, and (c) hard x-ray reflectivity (XRR) results compared with YXRO simulations using the STEM-determined and ideal structures. The dashed lines are intended to identify a few key peaks that are present in both experiment and simulations for the XRD data with the STEM structure. There is excellent agreement with the XRD data in (b), and for the YXRO simulations for the STEM structure in (c), with a slight shift of the calculated angle scale of ~0.06°, which can simply be the angular calibration in experiment or lack of full accuracy in the STEM-based model, the peak positions line up very well. The dip at 1.8° is no doubt reduced in expt. due to the more complex roughnesses in interfaces. The calculation for the ideal geometry again differs markedly from the experiment.

- **Soft x-ray reflectivity**

Figure S3(a) show the experimental soft x-ray reflectivity map as a function of photon energy and incidence angle for the SrO-terminated LSCO/LSMO multilayer. The reflectivity maps simulated using the STEM-determined structure and the ideal structure are shown together for comparison. The simulations of reflectivity were performed using the resonant index of refraction. The YXRO simulations using the STEM-determined structure show better agreement with the experimental data. To have a further comparison between experiment and theory, the experimental reflectivity at 931.2 eV is plotted in Fig. S4(a) and compared with the simulated reflectivities. Note the significant shift of the effective Bragg angle for the ideal geometry, from ~7.8° to ~7.3°, as expected because of its larger period thickness compared to the experimental ones. Figure S4(b) shows the experimental second derivative of the reflectivity data (black dots) and its comparison to the simulations using the STEM-determined structure. The simulations show good agreement with the experimental results, and demonstrate the sensitivity of the reflectivity to small deviations of the sample stacking layers from ideal. These are important considerations for future SW-RIXS and SW-XPS studies.

Soft x-ray reflectivity--Growth on SrO-terminated STO:

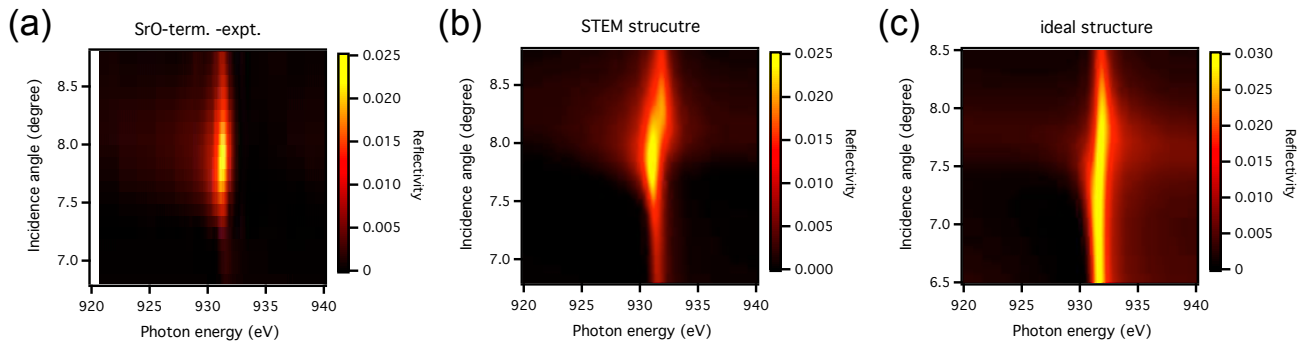

**FIG. S3.** Reflectivity maps as function of photon energy and incidence angle for the sample grown on TiO$_2$-terminated STO. (a) experimental data, (b) simulation with STEM-determined structure, as well as (c) ideal structure.

## Soft x-ray reflectivity--Growth on SrO-terminated STO:

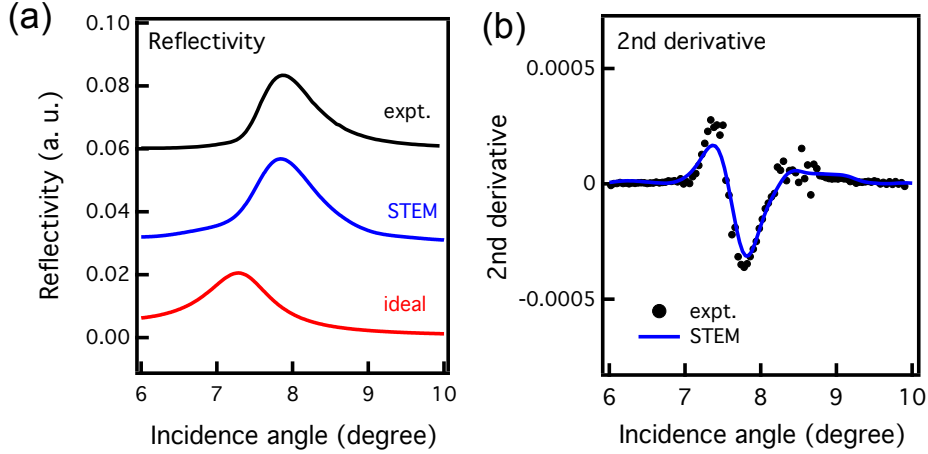

**FIG. S4.** (a) Experimental soft x-ray reflectivity as a function of incidence angle for the LSCO/LSMO multilayer grown on an SrO-terminated STO substrate, together with the simulated reflectivity using the STEM-determined (blue line) and ideal structure (red line). (b) 2$^{nd}$ derivative of the reflectivity data for SrO-terminated LSCO/LSMO multilayer (black dots) and the simulations using the STEM-determined structure (blue line) are plotted together for comparison.

We note here that the range 6°-10° encompasses the full range of significant reflectivity, and our experimental RIXS range in Figure 2 of the main text of 7.0°-9.5° captures most of this. In future work, it will be advisable to extend this range, but we do not feel that the present data range leads to any error in analysis.

## ➢ LSCO/LSMO multilayer grown on TiO$_2$-terminated STO--Structural information, SW-RIXS, and SW-XPS:

In this section, we present exactly the same SW-RIXS and structural data for the LSCO/LSMO multilayer grown on TiO$_2$-terminated STO, which, due to its irregular layer spacings, exhibits additional fine structure that makes the analysis less conclusive. In addition, we have obtained SW-XPS for this sample that supports the SW-RIXS analysis.

### • STEM, x-ray diffraction, and hard x-ray reflectivity

Figure S5 shows results for the TiO$_2$-terminated multilayer that are completely analogous to those in Figure S2 for SrO termination. Simple visual inspection of the STEM image shows much greater variation in the individual layer thicknesses, with these finally leading though the laterally-averaged red lines derived from STEM-HAADF to $d^{av}_{LSCO}$ = 24.8±2.7 Å, $d^{av}_{LSMO}$ = 28.5±5.0 Å and, in sum, $d^{av}_{ML}$ =

53.3±5.7 Å. Atomically resolved STEM electron energy loss spectra (EELS-not shown here) were used to determine the exact chemistry at the interfaces, each LSMO(top)/LSCO(bottom) and LSCO(top)/LSMO(bottom) interfaces which always consist of the sequence -La$_{0.9}$Sr$_{0.1}$O-La$_{0.9}$Sr$_{0.1}$O-CuO$_2$-La$_{0.66-x}$Sr$_{0.33+x}$O-MnO$_2$-La$_{0.66}$Sr$_{0.33}$O- and -La$_{0.66}$Sr$_{0.33}$O-MnO$_2$-La$_{0.9-x}$Sr$_{0.1+x}$O-CuO$_2$-La$_{0.9}$Sr$_{0.1}$O-La$_{0.9}$Sr$_{0.1}$O- respectively (0<x<0.15). These nos. are thus very close to those for the SrO-terminated sample, but with a much larger variation of the average LSMO thickness. Beyond this, comparing Figure S5(b) to Figure S2(b) shows that x-ray diffraction is significantly more complex for this sample, in particular showing spittings and distortions of all of the superlattices features. XRD calculations have again been optimized for the data for 20 layers divided as: 11 bilayers of 1.95 u.c. of $d_{LSCO}$ = 12.71 Å and 6.8 u.c. of $d_{LSMO}$ = 3.81 Å, which sum to $d_{ML}$ = 50.68 Å, 6 bilayers of 2.0 u.c. of $d_{LSCO}$ = 13.24 Å and 6.5 u.c. of $d_{LSMO}$ = 4.02 Å, which sum to $d_{ML}$ = 52.62 Å, and 3 bilayers of 2.0 u.c. of $d_{LSCO}$ = 13.24 Å and 6.8 u.c. of $d_{LSMO}$ = 4.06 Å, which sum to $d_{ML}$ = 54.10 Å. The more widespread splittings of the SL-N peaks are clearly predicted correctly, and overall spectra are in general agreement. Again, we conclude that our hard- and soft- x-ray calculations, as simulated with the YXRO program are much closer to the physics of the SW photoemission (or SW-XPS) data.

Comparing Figure S5(c) to Figure S2(c) near total reflection and the first-order multilayer Bragg reflection continues this complexity for TiO$_2$ termination, with experiment looking nothing like the ideal calculation with the YXRO program, but showing considerable agreement when this calculation is done with the detailed layer-by-layer STEM structure. We also expect this sample to show additional complexities in its SW profiles, beyond what a simple layer-by-layer average thickness can simulate, due to the lateral undulations and thickness variations present. Thus, even with the irregular layer-spacings of this sample, we should be able to semi-quantitatively describe its soft x-ray SW behavior.

TEM and x-ray reflectivity--Growth on TiO$_2$-terminated STO:

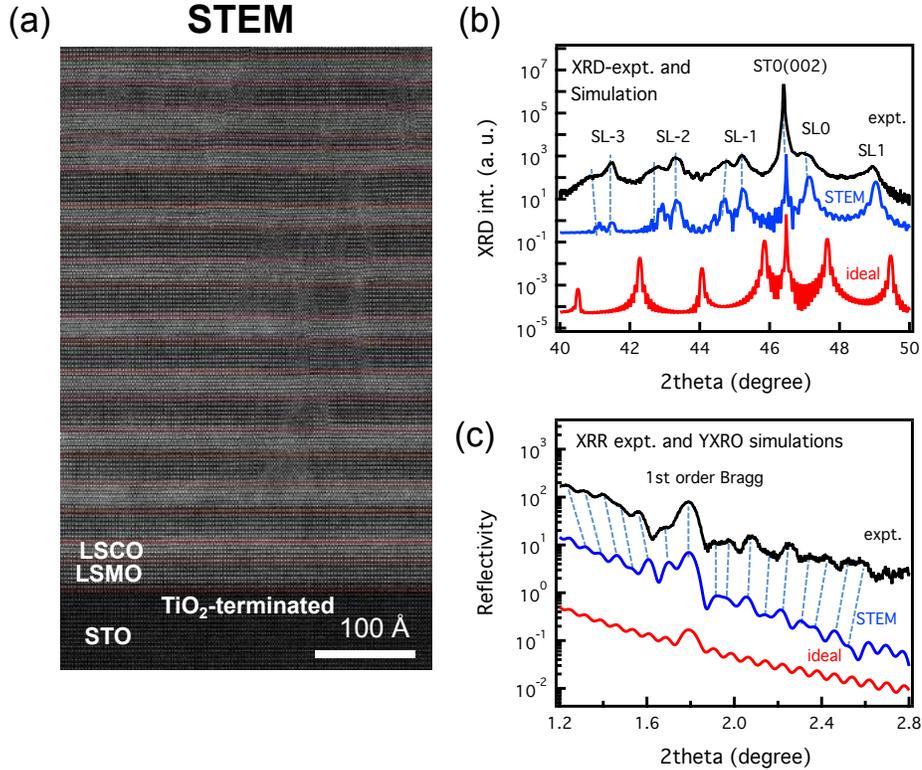

**FIG. S5.** Structure information for the LSCO/LSMO multilayer grown on the TiO$_2$-terminated SrTiO$_3$ substrate. (a) Scanning transmission electron microscopy (STEM) image, with red lines indicating lateral averages derived from STEM-HAADF data to determine the layer thickness variations through the sample. The experimental (b) (002) x-ray diffraction (XRD) and (c) hard x-ray reflectivity (XRR) results compared with the YXRO simulations using the ideal and STEM-determined structures. The dashed lines are intended to identify peaks that are present in both the experimental data and the simulations in (b) and (c). For this sample in (c), it appears that the experimental angle scale is ~16% larger than the STEM-based simulation, which can easily be explained by effects at low angle not fully incorporated in our laterally averaged structure.

- **Soft x-ray reflectivity**

Figure S6 is to be compared to Figure S3 for the SrO-terminated sample, and they actually show somewhat similar results. But the experimental Bragg peak for TiO$_2$ is somewhat broader, as might be expected, and the double-peaked structure in the STEM-based simulation is more pronounced in relative intensity, perhaps enhancing satellite peaks in the SW profile.

Soft x-ray reflectivity—Growth on TiO$_2$-terminated STO:

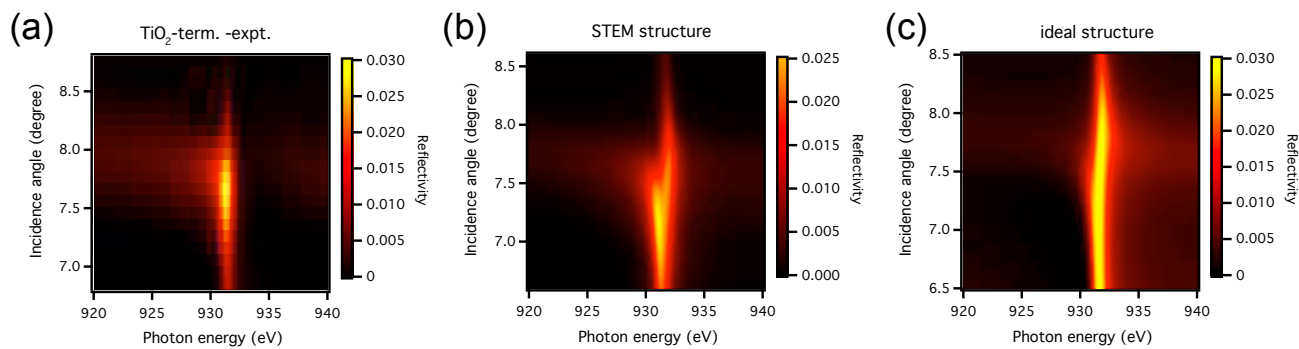

**FIG. S6.** Reflectivity maps as function of photon energy and incidence angle for the TiO$_2$-terminated sample (a) experimental data, (b) STEM determined structure, as well as (c) ideal structure.

Figure S7 of reflectivity at our energy of 931.2 eV for the TiO$_2$-grown sample can now be compared to Figure S4 for the SrO growth, and, although the straight reflectivity curves in Figure S7(a) may look similar, they are shifted by about 1° with respect to one another, and the those for TiO$_2$-grown are much wider on the low-angle side. Figure S7(b) of the second derivative also shows a satellite peak at about 8.4° that does not exist in Figure S4(b). This is another indicator that this sample will yield more complex SW-RIXS.

Soft x-ray reflectivity--Growth on TiO$_2$-terminated STO:

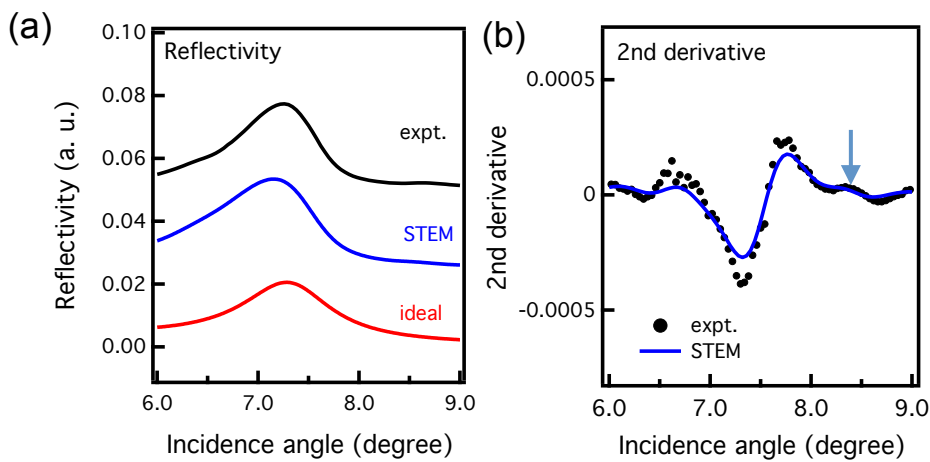

**FIG. S7.** (a) Experimental soft x-ray reflectivity as a function of incidence angle for the LSCO/LSMO multilayer grown on a TiO$_2$-terminated STO substrate, together with the simulated reflectivity using the STEM-determined (blue line) and ideal structure (red line). (b) 2$^{nd}$ derivative of the experimental reflectivity data (black dots) and the simulations using the STEM-determined structure (blue line) are plotted together for comparison. The blue arrow indicates a satellite peak not present for the SrO-grown sample.

● **Standing-Wave RIXS results**

Figure S8 showing SW-RIXS results for the $TiO_2$-grown sample can be compared to Figure 2 in the main text for the SrO-grown sample. Although the RIXS spectra and their deconvolution in Figure S8(a) are not surprisingly very similar to those in Figures 2(a),(b) the analogous curves resolved into the various losses in Figures S8(c),(d) are more complex, in particular showing satellite structure at higher angles. Simulations for various depth selections in the sample are again shown in these panels, and they do show shifts in the minima that are suggestive from panel (c) that the $d_{xy}$ and $d_{xz}/d_{yz}$ excitations again arise in the full bulk of the LSCO layer, but by contrast that the $d_{z^2}$ arises from the LSCO-top/LSMO-bottom interface. This indicates a different average interface structure in the $TiO_2$-grown sample, but its irregularity prevents quantifying this further. The simulations do not predict the satellite structure for higher angles in either of panels (c) or (d). Comparing experiment with simulations in panel (d) agrees qualitatively with Figure 2(d) in that the magnetic excitations are associated with the two types of interfaces, while the quasi-elastic intensity arises from mostly the bulk of the LSCO layer.

Thus, the $TiO_2$-grown sample is clearly more complex to analyze than the SrO-grown sample in the main text, but it nonetheless again exhibits SW-RIXS modulations that are at least qualitatively in agreement with the results for the SrO-grown sample. This sample thus further confirms SW-RIXS as a technique, but provides a warning in future experiments to make sure the multilayer structure is as highly regular as possible.

## SW-RIXS--Growth on TiO$_2$-terminated STO:

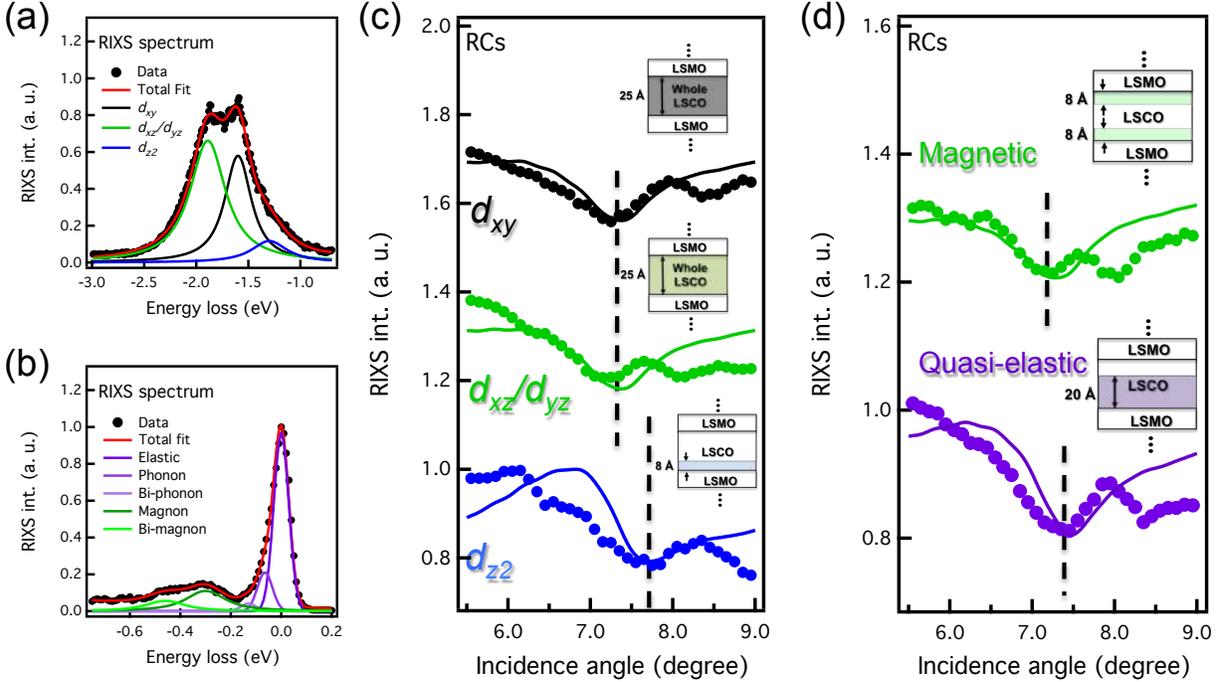

**FIG. S8.** SW-RIXS of various excitations for TiO$_2$-terminated sample. RIXS spectra of (a) *dd* excitations and (b) quasi-elastic and magnetic excitations. (c) The experimental *dd*-excitations RCs (data points) together with calculations (lines). (d) The experimental RCs for the magnetic and quasi-elastic excitations (data points) together with calculations (lines).

- **Standing-wave x-ray photoemission results**

Complementary standing-wave excited x-ray photoelectron spectroscopy (SW-XPS) measurements were also performed on the TiO$_2$-terminated sample to investigate the depth-resolved information from the LSCO and LSMO layers. Unfortunately, we do not have similar data for the SrO-grown sample. The incidence photon energy of the SW-XPS measurement was again 931.2 eV. Thus, the x-ray optics should be identical to that for the SW-RIXS measurements. Figures S9 (a) and (b) show the Cu 3p and Mn 3p core level spectra, with their corresponding RCs in (c). Since we are near the Cu L$_3$ absorption maximum, there are emergent Auger satellites near Cu 3p, but the RC for it has been derived from the lowest binding energy true photoemission peak. It is evident that the Cu 3p and Mn 3p RCs show markedly different shape and change of intensity due to the fact that the Cu and Mn photoelectrons arise from different layers. They also show weak satellites, especially on the high-angle side over 8.0-8.6° that are reminiscent of those in the SW-RIXS results from this sample in Figure S8. The results of the YXRO simulations, again for the STEM-derived structure, shown in Figure S9(d), are in excellent agreement with the experimental results, although do not show the satellites, except when they are amplified in the 2nd derivatives curves in Figure S9(e). These indicate likely fine structure at smaller and higher angles, and help in accepting the additional fine structures in the experimental data.

These SW-XPS results thus enhance our confidence in the analysis method used for both sets of SW-RIXS data, including the influence of non-regularity in the layer thicknesses of one of the samples studied.

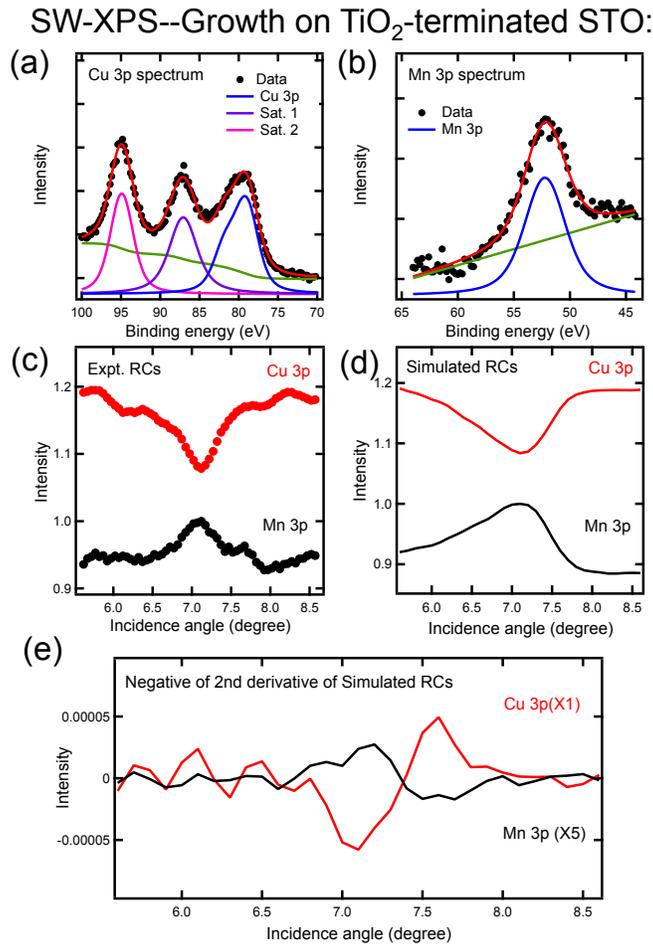

**FIG. S9.** Experimental SW-XPS results showing the depth-resolved information. (a) Cu 3p and (b) Mn 3p core level spectra for a $TiO_2$-terminated LSCO/LSMO multilayer. The core-level data (black dots) are fitted using Voigt functions and Shirley background (in green). (c) The photoelectron RCs of Cu 3p and Mn 3p. (d) The YXRO simulated RCs, including the negative second derivatives in (e) for both as well.

## Supplemental References